\documentclass[fleqn,10pt]{wlpeerj}
\usepackage{fixltx2e}
\usepackage{algorithm}
\usepackage{algorithmic}
\usepackage{array}
\usepackage{booktabs}
\usepackage{multirow}
\usepackage{amsmath}
\usepackage[utf8]{inputenc}
\usepackage[english]{babel}
\usepackage{graphicx}
\usepackage{float}
\usepackage{amssymb}
\usepackage[hyphens]{url}

\title{Predicting the Behavior of the Supreme Court of the United States: A General Approach}

\author[1]{Daniel Martin Katz}
\author[2]{Michael J Bommarito II}
\author[3]{Josh Blackman}
\affil[1]{Michigan State University - College of Law}
\affil[2]{Michigan State University - College of Law}
\affil[3]{South Texas College of Law}


\begin{abstract}
Building upon developments in theoretical and applied machine learning, as well as the efforts of various scholars including \cite{guimera2011justice}, \cite{ruger2004supreme}, and \cite{martin2004competing}, we construct a model designed to predict the voting behavior of the Supreme Court of the United States.  Using the extremely randomized tree method first proposed in \cite{geurts2006extremely}, a method similar to the random forest approach developed in \cite{breiman2001random}, as well as novel feature engineering, we predict more than sixty years of decisions by the Supreme Court of the United States (1953-2013). Using only data available prior to the date of decision, our model correctly identifies 69.7\% of the Court’s overall affirm / reverse decisions and correctly forecasts 70.9\% of the votes of individual justices across 7,700 cases and more than 68,000 justice votes.  Our performance is consistent with the general level of prediction offered by prior scholars.  However, our model is distinctive as it is the first robust, generalized, and fully predictive model of Supreme Court voting behavior offered to date.  Our model predicts six decades of behavior of thirty Justices appointed by thirteen Presidents.  With a more sound methodological foundation, our results represent a major advance for the science of quantitative legal prediction and portend a range of other potential applications, such as those described in \cite{katz2013quantitative}.
\end{abstract}

\begin{document}

\flushbottom
\maketitle
\thispagestyle{empty}

\section*{Introduction}

Each October, as the leaves begin to fall, the first Monday marks the beginning of another term of the Supreme Court of the United States.  Like the years past and the years that will follow, each term brings with it a series of challenging and substantively important cases covering a wide range of legal questions.  In a given year the Court might consider topics as diverse as tax law, freedom of speech, patent law, the law of searches and seizures, administrative law, due process, the proper scope of the takings clause, freedom of assembly, equal protection, environmental law, and many other legal questions.  In most instances, the Court’s decisions are meaningful not only for the litigants, but also for broader range of individuals, entities and social and political institutions. 
\\
\\
Predicting its decisions and attendant rationales is one of the great pastimes for observers of the Supreme Court. Every year, the authors of countless law reviews, journals, magazines, newspapers, television and radio pundits, blog posts, and tweets try to answer the questions that are on everyone's minds: How will the Court rule in a particular case? Will the Justices vote based on the political party of the President who appointed them? Will the Justices surprise us with an unexpected ruling?  In these and other related fora, individual commentators offer various theories about what the Court will do and why it will choose to do so.
\\
\\
The sheer number of qualitative explanatory and predictive theories is significant. As a matter of scientific forecasting, however, the quality of many of these theories is unclear.  It is hard to measure how a particular approach predicts earlier cases, and, more importantly, how accurately the model will predict future cases.  Without a robust, general model that has been subjected to proper validation, it is difficult to determine how any given forecasting method will perform over a period of time.  Most non-normative theories offered in traditional legal scholarship designed to forecast the Court’s behavior have not been tested in any manner that could demonstrate their forward predictive validity, either out-of-sample or forward prediction. That is, nearly all models have been designed ex post to review or explain cases already decided, rather than used \textit{ex ante} to predict future cases.  As noted in \cite{martin2004competing}, “the best test of an explanatory theory is its ability to predict future events. To the extent that scholars in both disciplines [social science and law] seek to explain court behavior, they ought to test their theories not only against cases already decided, but against future outcomes as well.”
\\
\\
Building on recent developments in theoretical and applied machine learning, as well as the efforts of various scholars including \cite{guimera2011justice}, \cite{ruger2004supreme}, and \cite{martin2004competing}, we construct a model designed to predict the voting behavior of the Supreme Court of the United States.  Using the extremely randomized tree method first proposed in \cite{geurts2006extremely}, a method similar to a random forest approach developed in \cite{breiman2001random}, as well as novel feature engineering, we apply our method to predict more than sixty years of decisions by the Supreme Court of the United States (1953-2013). Using only data available prior to the date of decision, our model correctly identifies 69.7\% of the Court’s overall affirm / reverse decisions and correctly forecasts 70.9\% of the votes of individual justices across 7,700 cases and more than 68,000 justice votes. Our performance is consistent with the general level of prediction offered by prior scholars.  However, our model is distinctive as it is the first robust, generalized, and fully predictive model of Supreme Court voting behavior offered to date.  Our model predicts six decades of behavior of thirty Justices appointed by thirteen Presidents.  With a more sound methodological foundation, our results represent a major advance for the science of quantitative legal prediction and portend a range of other potential applications, such as those described in \cite{katz2013quantitative}.

\section*{Predicting the Voting Behavior of the Supreme Court of the United States}

Every year thousands of petitioners appeal to have their case considered by the United States Supreme Court.  In most situations, the Court will decide to hear a case by a granting a petition for a \textit{writ of certiorari}.  If that petition is granted, each of the parties to the litigation will then submit written materials on the relevant issues and later provide oral argument before the Court.  Based upon the weight of the arguments before the Court and other factors, each participating justice (typically nine) ultimately cast his or her vote whether to affirm or reverse the decision of the lower court.  From a prediction standpoint, there are two related but distinct prediction questions:  \textit{(1) Will the Court affirm or reverse the lower Court’s judgment?}  \textit{(2) How will each individual Justice vote on the question before him or her?}  As these are typically questions considered in both academic and popular circles, these are the questions we seek to predict using the methodological approach we outline herein.

\subsection*{Properties of General Supreme Court Prediction Model}

There are at least three basic goals that should animate the construction of any Supreme Court prediction model or method.  As a general matter, a model should strive to be general, to be robust and to be fully predictive.  We are mindful of each of these goals when developing our method of Supreme Court decision-making.  

\subsubsection*{i. General}

Starting with the first of these goals, we aim to develop a method of Supreme Court prediction that is general.  Specifically, a method should not only work for a single year or short period but also should be \textit{generalized} such that it might work for any year.  This is important because it is unclear when making predictions \textit{ex ante} whether performance in a given year is systematic or the byproduct of unique features of a given docket or some other latent features.  In other words, it is impossible to know in advance if a given year will be similar to previous years, or significantly different. At a theoretical level, the effort to produce a general model can involve a tradeoff between local and global optimization whereby we must sacrifice some local performance at a specific time period for overall performance over a wide window of time. It is far easier to make accurate predictions for a given year and a known group of Justices, rather than to develop a model that works consistently over six decades.  
\\
\\
As the first major effort in this direction, more than ten years ago, political scientists and legal scholars (\cite{ruger2004supreme}) held a tournament that pitted expert predictors against a prediction algorithm based upon a classification tree.  The goal was straightforward – predict the votes of the individual justices as well as the ultimate decisions of the Supreme Court prior to the release of the Court's decision. For each case, a classification tree generated its predictions while the experts (law professors and practitioners) simultaneously submitted their selections.  With respect to predicting outcomes (i.e. affirm / reverse) for the 78 cases in the 2002-2003 term (the "October 2002 Term," as it is called), the prediction model correctly forecasted 75\% of the cases while the human experts correctly identified 59\%. For the votes of individual justices, the model was correct for 66.7\% of the justice votes while the human experts properly identified 67.9\%.
\\
\\
\cite{ruger2004supreme} represented a major contribution to the science of legal forecasting.  Their approach not only performed well in absolute terms but also matched or outperformed subject matter experts.  However, like all efforts it had important limitations.  Namely, their model was not general and it is now relatively clear that the methodological approach they employed is not well specified toward a general model of Supreme Court prediction. Namely, their approach was conducted during a ``natural court," which existed during one of the longest extended periods where there were no personnel changes on the Court, following Justice Stephen G. Breyer's appointment in 1994.  It is unclear how their model would perform in periods prior to 1994 or after 2005, following the replacements of Chief Justice William H. Rehnquist and Justices Sandra Day O'Connor, David H. Souter, and John Paul Stevens.   A general model would continue to offer accurate predictions, even with the appointment of different justices and across a wide range of social, economic and political contexts.  

\subsubsection*{ii. Robust}
While there is a reasonable level of temporal stability in the decisions of the Court, one cannot know in advance whether the upcoming Court will behave in a manner similar to its predecessors.  As such, the potential for overfitting remains ever present.  Over wider windows of time, various factors such as justice retirements, justice and court level ideological shifts, doctrinal shifts, a changing docket composition as well as a changing macro-level social, economic and political environment likely all impact the task of judicial prediction.  The ultimate performance of a prediction is a function of the system variability and the overall diversity of the case space.  Therefore, it is important to select a method that is known to be \textit{robust}. That is, it tends to neither overfit nor underfit the respective data. 
\\
\\
While the results of the \cite{ruger2004supreme} tournament were very promising, in the decade that followed this initial tournament, there was little subsequent Supreme Court prediction scholarship.  Other than a few notable exceptions such as \cite{blackman2011fantasyscotus} and \cite{guimera2011justice}, there were few extensions or improvements of this initial approach.   In reviewing \cite{ruger2004supreme}, it appears that the classification method undertaken was not well suited to the task of robust generalization. Specifically, in an effort to support model transparency for a non-technical audience (i.e. lawyers and law professors) the authors used a standard single classification tree in order to generate their predictions. Specifically, they note “[O]ur choice to use classification trees is motivated by the transparency of the model; i.e., trees are produced that can be graphically represented and easily studied.”  
\\
\\
One well-known problem with the standard classification tree approach applied in \cite{ruger2004supreme} is its tendency toward overfitting.  As noted by \cite{hastie2009elements}, a single classification tree is notoriously noisy.  Individual classification trees typically feature high levels of variance.  As such, the variance of the final estimator is typically very high.  Thus, results are sensitive to small changes in the data and can even significantly vary across various model runs.  Several developments in the theoretical machine learning literature partially alleviate this issue.  Specifically, in the years following \cite{breiman1984classification}, significant work on classification and regression trees (CART) has improved their general performance and helped avoid the problem of overfitting. \cite{breiman2001random} refines \cite{breiman1984classification} and offers a substantial modification of his prior work on \textit{bagging} “by building a large collection of de-correlated trees, and then averaging them.”  Rather than relying on a single decision tree for a small period of time, the random forest approach and other related ensemble methods are designed to rely on many randomized decision trees in order to reduce the variance of the respective predictive estimator.  Given their relative simplicity, random forests and other related methods such as extremely randomized trees, have proven to be highly effective on real world data. 

\subsubsection*{iii. Fully Predictive}
In addition to being general and robust, any complete model of Supreme Court prediction should also be \textit{fully predictive}.  In an important recent paper on Supreme Court prediction, \cite{guimera2011justice} engage a very particular form of prediction of the Court’s behavior from 1953-2005. They note "[W]e want to predict the vote of a justice in case \textit{n} (without loss of generality we set this justice to be number 1), given the complete voting record of the court up to case \textit{n - 1} and the votes of the other eight justices in case \textit{n}."  In other words, the authors are predicting the ninth row/column of a matrix given the values of the other eight rows/columns.  Or stated differently, given the votes of all Justices in all previous cases, \textit{and the votes of the eight other Justices in the current case}, their model forecasts the vote of the ninth Justice in the current case. While this approach provides some important insights about the nature of inter-justice voting patterns, it is not a fully predictive model like that undertaken in \cite{ruger2004supreme}.  It does not allow the prediction of how all nine Justices, currently on the Court, will vote.
\\
\\
At best, it can be characterized as a partially predictive model of Supreme Court behavior.  The goal of Supreme Court prediction faced by scholars and litigators is to forecast an individual justice as well as the overall Court’s decision without any prior knowledge of the votes of his or her fellow justices.  Thus, a complete model should rely upon a similar information set and undertake the same substantive task that is typically faced by human subject matter experts.   

\subsubsection*{iv. Toward a General, Robust and Fully Predictive Method}

In sum, none of the existing approaches applied to question of Supreme Court prediction simultaneously achieve the three basic goals of being general, robust and fully predictive. \cite{ruger2004supreme} is fully predictive but unfortunately, their approach is not general and their method is not robust. By contrast, \cite{guimera2011justice} is general and robust, however, it is not  fully predictive.  As described below, our approach is the first offering in the literature that satistifes all three of these important criteria and thus represents a significant advance in the science of quantitative legal prediction.   

\section*{Trees, Forests and Extremely Random Trees: Advances in the Science of Binary Prediction}

The applied case of Supreme Court prediction presents a problem of binary classification where we seek to forecast whether an individual justice, as well as the Supreme Court as a whole, will decide to either \textit{affirm} or \textit{reverse} the judgment of a lower court.   In this instance of binary classification, we take the matrix of observable prior data to learn / construct a prediction function. Using that function, we evaluate a future case and try to predict its corresponding outcome (i.e. affirm or reverse). In our taxonomy, \textit{y=0} if the Court, or a Justice, will affirm the lower court’s decision. \textit{y=1} if the Court, or a Justice, will reverse the lower court’s decision. We use the individual justice vote predictions to forecast the overall decision of the Supreme Court.  
\\
\\
A wide variety of supervised machine learning methods have been developed to learn or recover the best performing function \textit{f(x)}.  One popular approach commonly used for binary classification problems is the classification and regression trees (CART) methods first offered in \cite{breiman1984classification}. The standard CART approach was the method relied on by \cite{ruger2004supreme}. Using a single decision tree, the authors forecast the respective votes of Supreme Court justices for the October 2002 Term.  As noted earlier, the performance of the tree outperformed human experts, including notable law professors and well-regarded appellate lawyers.  Of important note, in the only tournament ever conducted using highly renowned lawyers, the experts properly classified 59\% of the Court’s overall affirm / reverse rate correctly, while properly forecasting 67.9\% of the votes of individual Justices.  In terms of case level prediction, the accuracy rate for the experts was only slightly better than chance (\cite{blackman2011fantasyscotus}). The classification tree, by contrast, correctly forecasted 75\% of the case level outcomes and 66.7\% of the Justice level votes.  The performance of this group of experts is illustrative of the difficulty of the prediction problem.  While a small number of the Court’s decisions are relatively straightforward, the full docket--on average about 80 cases per year--contains a number of complex questions that are challenging to forecast even for well-respected Supreme Court experts.
\\
\\
The individual trees generated using the CART approach are high variance objects. As a function of the random perturbations of the model an individual tree may underfit or overfit the data.  Building upon the work of \cite{ho1998random}, \cite{breiman2001random} proposed an ensemble method that leverages two forms of randomness - the \cite{breiman1996bagging} idea of \textit{bagging} (bootstrap aggregation) with \textit{random substrates}.  One simple manner in which to understand random forests, extremely randomized trees and other related ensemble methods is to consider the ideas underlying the wisdom of the crowds (\cite{surowiecki2005wisdom}, \cite{page2008difference}, 
\cite{fisher2009perfect}, \cite{rauhut2011wisdom}). Ensemble methods leverage the wisdom of the statistical crowds by generating a number of diverse trees and then averaging across the entire forest. While an individual statistical learner (a single tree) might offer an unrepresentative prediction of a given phenomena, under certain conditions, the crowdsourced average of a larger group of learners is often better able to forecast various sets of outcomes.  By generating many different decision trees and then averaging over the results, ensemble methods can convert a set of otherwise weaker learners into a collectively strong learner.
\\
\\
As highlighted in important investigations such as \cite{caruana2006empirical}, \cite{diaz2006gene} and \cite{caruana2008empirical} ensemble methods have proven to be highly effective for a large number of classification tasks. In addition, through  \cite{caruana2006empirical} and \cite{caruana2008empirical} random forests and related methods have been shown to be more robust than both the standard (CART) approach and a wide variety of other competing methods. 
\\
\\
Ensemble methods have been applied by various scholars in a wide variety of contexts (e.g.  \cite{irrthum2010inferring}, \cite{moosmann2008randomized}, \cite{diaz2006gene}, \cite{shi2004tumor}, \cite{svetnik2003random}). In the years following \cite{breiman2001random}, several alternative but allied methods have been developed.  Building upon the highly influential work on random forests offered in \cite{breiman2001random}, \cite{geurts2006extremely} offers an alternative ensemble which embeds even more randomness into the tree construction process.  Namely, ``[W]ith extremely randomized trees, randomness goes one step further in the way splits are computed. As in random forests, a random subset of candidate features is used, but instead of looking for the most discriminative thresholds, thresholds are drawn at random for each candidate feature and the best of these randomly-generated thresholds is picked as the splitting rule. This usually allows to reduce the variance of the model a bit more, at the expense of a slightly greater increase in bias."\footnote{See ``1.9.1.2. Extremely Randomized Trees" available at http://scikit-learn.org/stable/modules/ensemble.html}  In order to remain robust against the temporal changes present in the Court's behavior over the past sixty years, we generate a set of extremely randomized trees (extra-trees) in the analysis that follows.  
\\
\\
\section*{Data and Variables}

In order to build our prediction model, we rely on data from the Supreme Court Database (SCDB).  The SCDB features more than six decades of high-quality expertly coded data on the Court’s behavior.\footnote{Harold J. Spaeth, Sara Benesh, Lee Epstein, Andrew D. Martin, Jeffrey A. Segal, and Theodore J. Ruger. 2013. Supreme Court Database, Version 2013 Release 01. URL: http://supremecourtdatabase.org. Last accessed: July 5, 2014.}  A product of years of dedication by Professor Harold Spaeth as well as others, the database has been consistently subjected to reliability analyses and has been used in hundreds, if not thousands of academic studies (e.g. \cite{segal2002supreme}, \cite{bailey2008does}, \cite{benjamin2012standing}, \cite{epstein2007ideological}, 
\cite{segal1996influence}).  While there are important limits (i.e. \cite{shapiro2008coding}), the SCDB features up to two hundred and forty seven variables for each case including background variables, chronological variables, substantive variables, outcome variables, voting variables and opinion variables. 
\\
\\
We define the outcome of our prediction variable based on the decision of the lower court. We consider whether in a given case \textit{n} the Supreme Court of the United States will either affirm or reverse the decision of the lower court.  Thus, our left hand side variable is the binary outcome variable \textit{0= affirm, 1= reverse}.  As noted earlier, for cases with multiple issues (the Court affirms in part, reverses in part) we consider the primary issue as set forth in SCDB.  In addition, we exclude decisions that do not contain actual justice votes (such as \textit{per curiam} decisions). 
\\
\\
In order to predict our dependent variable, we leverage a wide number of features that have been previously shown to be meaningful in the existing explanatory theories of Supreme Court decision-making. The full list of these variables is offered in \textit{Figure 1}.  In the feature selection process, we employ a mixture of variables including those drawn directly from the Supreme Court Database denoted as [S], the Segal-Cover Scores [SC] and those we developed through various feature engineering [FE].  Each time the model is retrained, we allow the learner to explore the space and identify the optimal configuration that best predicts the Court's behavior based on the large number of features presented in \textit{Figure 1}.

\begin{figure}
\centering
\includegraphics[width=\linewidth]{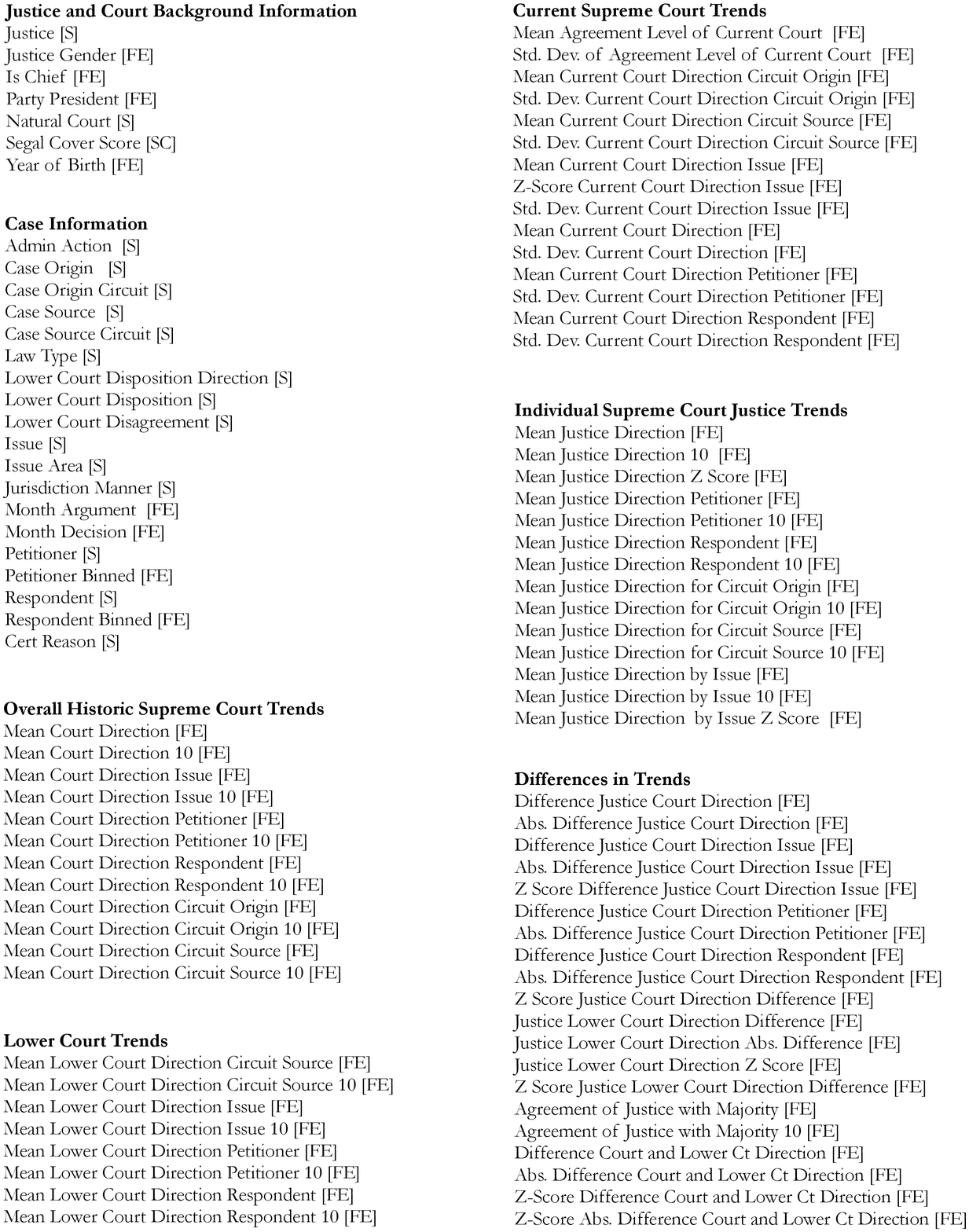}
\caption{Variables Employed by the Model}
\label{Variables Employed by the Model}
\end{figure}

\subsubsection*{i. Court and Justice Level Information} 

As displayed in \textit{Figure 1}, we include several forms of variables in our model. These include court level and justice-level variables such as \textit{party of appointing president}, \textit{segal-cover nomination score}, \textit{year of birth} and \textit{natural court} (the Court era in which the decision was authored).  Along with other variables, the first two features capture some of the ideological dimensions that explain at least some of the Court’s decisions.  The \textit{year of birth} and \textit{natural court} are control variables that we believe might capture an underlying time dynamic or age dynamic in the vote space.

\subsubsection*{ii. Case Information}

In addition, we include various case-centric features that we have some theoretical basis for believing might impact the propensity of various judges to vote in a certain manner.   As coded by Spaeth, these include the \textit{issue}, \textit{issueArea}, \textit{lawType}, \textit{certReason}, \textit{respondent}, \textit{petitioner}, \textit{caseOrigin}, \textit{caseSource} and \textit{lcDispositionDirection}.  In addition, we conduct some basic feature engineering that allows us to consider the \textit{monthArgument} and \textit{monthDecision} (which tracks the current month of the model).  At first glance, this last variable would appear to be using out of sample information, as this information is only known after the case is decided.  However, our reported predictions are derived from predictions generated each day until the day of decision.  Thus, including this variable still places us on equal footing with any human predictor who might attempt to forecast the Court’s decision, and later decide to shift the forecast at some point prior to the Court's decision.  Given the granularity of the respondent and petitioner variables (i.e., more than 300 discrete values), we aggregate the relevant categories into higher order bins under the general belief that it might better aid prediction.  These are labeled as \textit{respondentbinned} and \textit{petitionerbinned}, respectfully. 

\subsubsection*{iii. Historical Justice \& Court Information}

Historical information forms the basis for all predictions.  In order to aid our prediction model, we engage in some basic feature engineering  regarding baseline trends in the Court’s behavior.  As new Justices have joined, the court’s average ideology has shifted over the years (\cite{martin2002dynamic}, \cite{bailey2013}). Thus, variables such as \textit{courtdirectionmean} and \textit{courtdirectionstd} are designed to track the average and standard deviation in the Court’s historical behavior on an ideological scale of (0,1) where 0=max liberal and 1=max conservative.  
\\
\\
An additional source of temporal variation in the Court’s behavior is the ideological shift of various individual justices (\cite{epstein2007ideological}, \cite{martin2007assessing}, \cite{epstein2007perils}).  Thus, through the \textit{justicedirectionmean}, and \textit{justicedirectionstd} features we account for any time varying shift in the average ideology of individual justices.  The \textit{justicecourtdifference-z} variable tracks differences as measured in z-scores between an individual justice and the overall court.   
\\
\\
Our method is designed to identify and leverage features that contribute to prediction.  Thus, we are somewhat over-inclusive with respect to adding model features.\footnote{Our complete dataset and associated code is all publicly available on Github: https://github.com/mjbommar/scotus-predict}  As displayed in \textit{Figure 1}, we include more than ninety total variables in our model.  Taken together, we believe that all of these variables might meaningfully contribute to predicting individual justices as well as the overall Court’s behavior in the cases they collectively consider.     

\section*{Methods}
\subsection*{Formalizing a Set of Extremely Randomized SCOTUS Trees}

The search for the optimal tree configuration is an \textit{NP-complete problem} (\cite{hyafil1976constructing}). Thus, the optimal solution to such tree configuration problems cannot be determined in advance.  For anything other than a trivial problem, this implies that we must rely upon some sort of heuristic solution in the tree construction process.  Luckily, there exist a number of well performing heuristics designed to solve for the optimal tree configuration using various approximate solution methods (e.g. \cite{murthy1998automatic}, \cite{chou1991optimal}, \cite{safavian1991survey}).  
\\
\\
In recent years, there have been major advances in classification tree methods. In empirical investigations such as \cite{caruana2006empirical} and \cite{caruana2008empirical}, ensemble methods have been shown to be unreasonably effective on real world data.  Ensemble methods such as random forests, extremely randomized trees and fully random trees leverage several forms of randomness – (1) randomness in the data and (2) randomness in the features (potential splits).   For example, the \cite{breiman2001random} random forest method applies bootstrapping aggregation to the row of our training data while also sampling random substrates of the variables listed in the columns. A final ensemble create using such an approach is thus theoretically grounded and typically produces a robust method prediction with much lower variance than that produced using the CART approach. 
\\
\\
A close analog of random forests are extremely randomized trees (ERT). We apply ERTs (\cite{geurts2006extremely}) in our analysis.  Despite its similarity to random forests, there are important differences.  First, extremely randomized trees do not rely on the bagging procedure as outlined in \cite{breiman1996bagging}.  Instead, the same input training set is used to train all of the trees in question (\cite{geurts2006extremely}). In addition, ERT selects the split by indexing randomly across both the variable index and variable splitting value.  By contrast, the random forest select the optimal splitting condition among a random subset of variables.  As noted in (\cite{geurts2006extremely}), ``[F]rom the bias-variance point of view, the rationale behind the Extra-Trees (extremely randomized trees) method is that the explicit randomization of the cut-point and attribute combined with ensemble averaging should be able to reduce variance more strongly than the weaker randomization schemes used by other methods. The usage of the full original learning sample rather than bootstrap replicas is motivated in order to minimize bias."
\\
\\
The full pseudo-code of the extremely randomized trees (extra-trees) algorithm as outlined in \cite{geurts2006extremely} is offered below:
\newpage

\begin{algorithm}[H] 
\textit{}
\floatname{algorithm}{}  
\renewcommand{\thealgorithm}{Pseudo-Code of the Extra Trees Algorithm}  
    \caption{- Geurts, et al. (2006)}
\vspace{1.75 mm}
\textbf{Build\_an\_extra\_tree\_ensemble}(\textit{S}). 
\\
\textit{Input}: a training set \textit{S}. 
\\
\textit{Output}: a tree ensemble ${T}$ = \{$t_1$,..., $t_M$\}.
\\
--For \textit{i}=1 to \textit{M} 
\\
\textbullet\ Generate a tree: $t_i$=\textbf{Build\_an\_extra\_tree}(\textit{S});
\\
--Return ${T}$.
\\
\\
\textbf{Build\_an\_extra\_tree}(\textit{S}). 
\\
\textit{Input}: a training set \textit{S}. 
\\
\textit{Output}: a tree ${t}$
\\
--Return a leaf labeled by class frequencies in \textit{S} if 
\\
(i) $|\textit{S}|$ $<$ $n_\textit{min}$, or
\\
(ii) all candidate attributes are constant in \textit{S}, or 
\\
(iii) the output variable in constant in \textit{S}
\\
--Otherwise:
\\
1. Select randomly \textit{K} attributes, \{$a_1, ...,a_K$\}, without replacement, among all (non constant in \textit{S}) candidate attributes; 
\\
2. Generate \textit{K} splits \{$s_1$, ...,$s_K$\}, where $s_i$ = \textbf{Pick\_a\_random\_split}(\textit{S}, $a_i$), $\forall$\textit{i} = 1, ..., \textit{K};
\\
3. Select a split $s_*$ such that Score($s_*$, \textit{S}) = $max_{i=1,...\textit{K}}$ Score($s_*$, \textit{S})
\\
4. Split S into subsets $\textit{S}_l$ and $\textit{S}_r$ according to the test $s_*$;
\\
5. Build $t_l$ = \textbf{Build\_an\_extra\_tree}(\textit{$S_l$}) and $t_r$ = \textbf{Build\_an\_extra\_tree}(\textit{$S_r$}) from these subsets;
\\
6. Create a node with the split $s_*$, attach $t_l$ and $t_r$ as left and right subtrees of this node and return the resulting tree \textit{t}.
\\
\\
\textbf{Pick\_a\_random\_split}(\textit{S},\textit{a})
\\
\textit{Input}: a training set \textit{S} and an attribute \textit{a}.
\\
\textit{Output}: a split.
\\
--If the attribute \textit{a} is numerical:
\\
\textbullet\ Compute the maximal and minimal value of \textit{a} in \textit{S}, denoted respectively by $\textit{a}^\textit{S}_\textit{min}$ and $\textit{a}^\textit{S}_\textit{max}$;
\\
\textbullet\ Draw a cut-point $a_\textit{c}$ uniformly in [$\textit{a}^\textit{S}_\textit{min}$, $\textit{a}^\textit{S}_\textit{max}$];
\\
\textbullet\ Return the split [a $<$ $a_c$].
\\
--If the attribute \textit{a} is categorical (denote by $\mathnormal{A}$ its set of possible values):
\\
\textbullet\ Compute $\mathnormal{A}_\textit{S}$ the subset of $\mathnormal{A}$ of values of \textit{a} that appear in \textit{S};
\\
\textbullet\ Randomly draw a proper non empty subset $\mathnormal{A}_\textit{1}$ of $\mathnormal{A}_\textit{S}$ and a subset $\mathnormal{A}_\textit{2}$ of $\mathnormal{A}$ $\backslash$$\mathnormal{A}_\textit{S}$;
\\
\textbullet\ Return the split [\textit{a} $\in$ $\mathnormal{A}_\textit{1}$ $\cup$ $\mathnormal{A}_\textit{2}$].

\end{algorithm}   
\vspace{4.25 mm}
Using the extra-trees algorithm, we store a time-ordered subset of the Supreme Court Database (SCDB) and other derived features in a feature matrix for all cases prior to the current case. This includes data for each case and each justice indexed up to the \textit{n - 1} case.  From this feature matrix, we derive the individual trees and overall ensemble following the protocol outlined in the pseudocode. We apply the default settings of the ExtraTreesClassifier (Extremely Randomized Trees) with limited exceptions.\footnote{See \cite{scikit_learn}, in particular, 3.2.3.3.3. sklearn.ensemble.ExtraTreesClassifier \\http://scikit-learn.org/stable/modules/generated/sklearn.ensemble.ExtraTreesClassifier.html.  The parameters used to train the classifier are as follows:  'classify\_min\_samples\_leaf': 2, 'classify\_max\_depth': 32,'classify\_max\_features': 24, 'classify\_n\_estimators': 4000.} Given the data available up to the \textit{n - 1} case (the last case decided before the case we are attempting to predict), we apply the latest instance of our extremely randomized tree ensemble.  In other words, using the derived ERT, we pass the justice, case and overall court level features for the current case to the current set of extremely randomized trees and output a prediction each Justice. Then using this set of justice level forecasts, we can then construct a case level prediction using majority rule. 
\\
\\
In order to validate our model, we apply stratified k-fold cross-validation with 10 folds per training.  At each training step, we divide our training data into ten samples.  We then train 9 models on each of the nine subsets of nine folds, testing these models on the remaining ``holdout'' sample.  We select the model that performs best on the holdout sample as determined by its $F_1$ score.  Using the resulting final model at each training step, we can predict how each individual justice, and therefore the entire Court, will vote for a specific case.\footnote{Additional details regarding our model implementation is publicly available on Github - \\ {https://github.com/mjbommar/scotus-predict}}  This determination is based on all previous decisions for that Justice, the Court, and all previous cases.  This mimics the type of methodology that a Supreme Court expert would rely on when making their own \textit{ex ante} predictions.  
\\
\\
\\
\section*{Results}
The Supreme Court Database offers coverage from 1946-2013 (the end of the most recent Supreme Court term).  We train our model on the period from 1946 -1953, under the leadership of Chief Jusice Fred Vinson (known as the Vinson Court). We begin making forward prediction starting with the first case of the Warren Court in 1953, through the end of the 2012-2013 term.  For each of the predictions, offered over 60 years--7,700 cases and in excess of 68,000 individual justice votes--we only rely on data that would have been available prior to the Court’s decision.  We depart from \cite{guimera2011justice} by not leveraging information about the votes of other Justices in the current case in order predict any other justice's votes.  In other words, we construct a fully predictive model, which relies entirely on information available prior to the decision of the Court. No future information is relied on.
\\
\\
We restrict our analysis to cases with an actual written decision and assigned justice votes.  Thus, our analysis does not include \textit{per curiam} decisions or decisions that were dismissed on procedural grounds.  Following the convention used by Harold Spaeth and his successors, we seek to predict the cases’ primary issue dimension as defined in the respective database.  This follows the basic logic of issue-based voting outlined in the formal modeling literature by \cite{anderson2007}.  Although we could in principle disaggregate any cases respective of its issue dimensions and feed them into the model, for purposes of consistency with prior scholars, we restrict our analysis to standard cases.
\begin{figure*}[h]
\centering
\includegraphics[width=\linewidth]{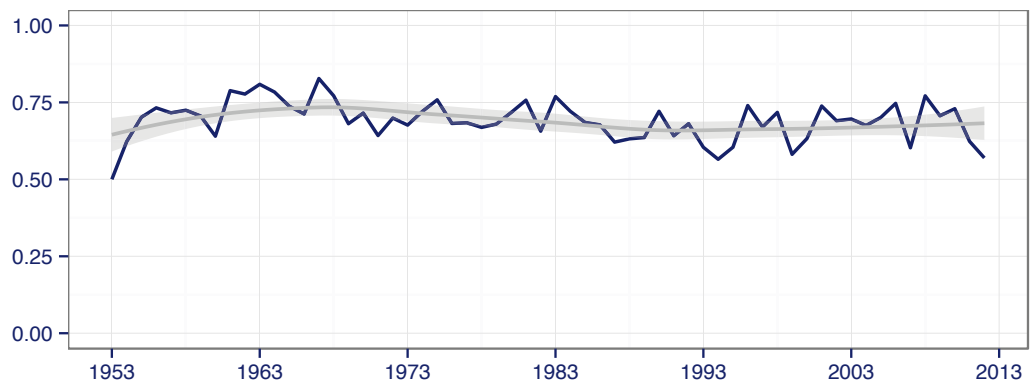}
\caption{Annual Accuracy of SCOTUS Extreme Randomized Trees}
\label{testing}
\end{figure*}
\\
\\
There are two forms of prediction that we undertake in this study – (1) justice level vote prediction, and (2) the prediction of the overall case outcome. Applying the extremely randomized trees approach to each case from 1953-2013, our model correctly forecasts 69.7\% of Case Outcomes and 70.9\% of Justice Level Vote Outcomes over the sixty year period.  While the results are close, somewhat surprisingly our model performs slightly better for the justice votes than on the cases themselves. We believe that this partly due to our success at forecasting the behavior of those distant from the Court's ideological center. 
\\
\\
Seen from a year over year perspective, there is a reasonable amount of variability in the model performance.  This is likely a function of changing predictability of the Supreme Court case space.  It is also likely a function of the inherent complexity of the forecasting problem.  The intermittent variability of our performance, however, does not provide the whole story. To provide a more complete perspective, \textit{Figure 2} plots both the yearly performance (blue) our model together with local polynomial regression (loess) (\cite{cleveland1988locally}). This loess regression fit provides a perspective on the performance of our model over time.  While the performance of our model varies from year to year, the trend remains relatively consistent over the sixty year period. 
\\
\\
Importantly, our method performs against a backdrop of justice transitions, shifting size and composition of the docket, as well as different macro political and economic conditions.  Our results compare favorably to those presented in prior work (\cite{guimera2011justice}, \cite{ruger2004supreme}) but are based upon the first generalized, robust and fully predictive model offered to date.

\subsection*{Exploring the Confusion Matrices and the Asymmetry in Affirm / Reversal Rates}

The case level confusion matrix displayed in \textit{Table 1} highlights the case level performance of the overall model from 1953-2013. The primary diagonal represents the true positive (we predicted an affirm, and the Court affirmed) and true negative (we predicted a reverse, and the Court reversed) elements. The off-diagonal elements are the false positive (we predicted an affirm, and the Court reversed) and false negatives (we predicted a reverse, and the Court affirmed).  A review of the matrix reveals an important property of the Court’s decision making.  Namely, the Court reverses a majority of the cases it accepts for review. As displayed in \textit{Figure 3} below, over the past sixty years, it is quite common for the Court to reverse in excess of 60\% of the cases it considers in any given year.  Our model properly identifies a significant percentage of these reversal decisions correctly.  
\vspace{5 mm}
\begin{table}[h]
\centering
\setlength{\arrayrulewidth}{.5mm}
\setlength{\tabcolsep}{14pt}
\renewcommand{\arraystretch}{1.3}
\begin{tabular}{ |p{2.25cm}|p{2cm}|p{2.25cm}||p{1.5cm}| }
\hline
 & \small Predict Affirm & \small Predict Reverse & \small Total\\
\hline
\small Court Affirms & 767 &1,856 &2,623 \\
\small Court Reverses & 474  & 4,603 & 5,077 \\
\hline
\small Total &1,242 & 6,458 &7,700 \\
\hline
\end{tabular} 
\caption{\label{tab:Table 1}Case Level Confusion Matrix}
\end{table}
\\
\\
By contrast, false positives, where we predict an affirm and the Court reverses, drive a significant percentage of the error in our model. The performance of our Supreme Court extremely randomized trees approach suffers most on the difficult task of determining when the Court will affirm the decision of the lower court (false positives). On average, when the Court grants certiorari, it does so to reverse a lower court decision, not to affirm it.
\\
\\
In 1,856 of the cases, the model incorrectly predicts that the court will reverse when it actually affirmed the lower court’s decision.  These false positives are likely generated in part from the underlying asymmetry in the Court's affirm / reversal rates as displayed in \textit{Table 1} and \textit{Figure 2}. Our model also likely struggles with the time varying nature of this asymmetry. Namely, the reversal percentage shifts from year to year and is, of course, not guaranteed to remain consistent going forward.   Consider \textit{Figure 3}, which displays the annual percentage of case reversed from 1953-2013.  The percentage fluctuates widely from as high as 77.5\% to as little as 50\%.  In the most recent years, since the start of Chief Justice Roberts's tenure, the percentages of cases that are reversed have dipped but thus far remain well within historic averages.  
\\
\begin{figure*}[h]
\centering
\includegraphics[width=\linewidth]{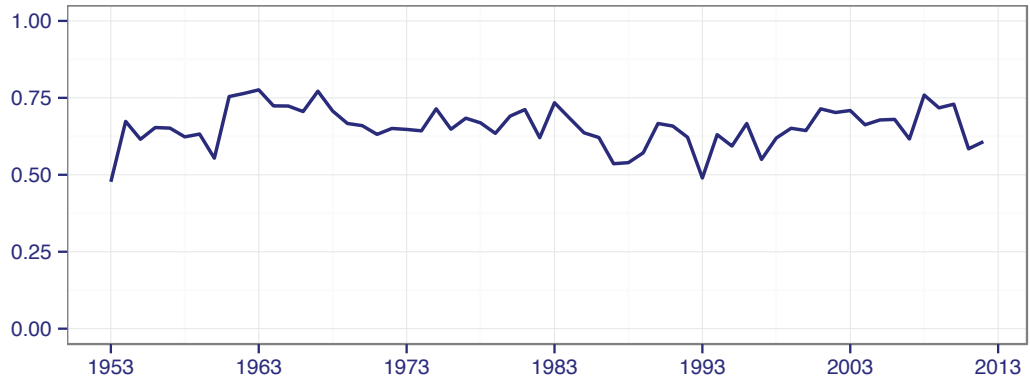}
\caption{Annual Case Overturn Percentage}
\label{testing}
\end{figure*}

The justice vote confusion matrix displays a similar pattern to the case confusion matrix.  In terms of overall error, we generate roughly four and a half times as much total error from instances where we predict a reversal and the court affirms the lower court (false negative) than instances where we predict affirm and the court reverses (false positive).   In other words, false positives far outstrip false negatives. In terms of precision and recall, however, the results are far more balanced. 

\vspace{2 mm}
\begin{table*}[h]
\centering
\setlength{\arrayrulewidth}{.3mm}
\setlength{\tabcolsep}{14pt}
\renewcommand{\arraystretch}{1.3}
\begin{tabular}{ |p{2.25cm}|p{2cm}|p{2.25cm}||p{1.5cm}| }
\hline
 & \small Predict Affirm & \small Predict Reverse & \small Total\\
\hline
\small Justice Affirms & 11,751 & 13,816 & 22,567 \\
\small Justice Reverses & 6,233  & 37,164 & 43,397 \\
\hline
\small Total &17,984 & 50,980 &68,964 \\
\hline
\end{tabular} 
\caption{\label{tab:Table 1}Justice Vote Confusion Matrix}
\end{table*}

\section*{The Relative Contribution of Particular Features to the Overall Prediction}
What features actually contribute to forecasting the behavior of the Supreme Court?  What is the relative contribution of these factors to overall prediction? There exists a long-standing debate about the subset of the feature space that actually assists in predicting judicial decision making. Traditional legal scholars tend to emphasize the legal features and legal questions presented in individual cases.  They tend to downplay weighted non-legal factors such as judicial ideology and its contribution to the shape of the law, and focus on jurisprudence and formal legal doctrine.  Legal realists and social scientists who study judicial behavior have demonstrated the incompleteness of this traditional description of the Court decision-making.  While there is likely merit in elements of many existing theories, the challenging question is how to properly characterize the ensemble of legal, political and social factors which collectively drive observed outcomes. 
\\
\begin{figure}
\centering
\includegraphics[width=\linewidth]{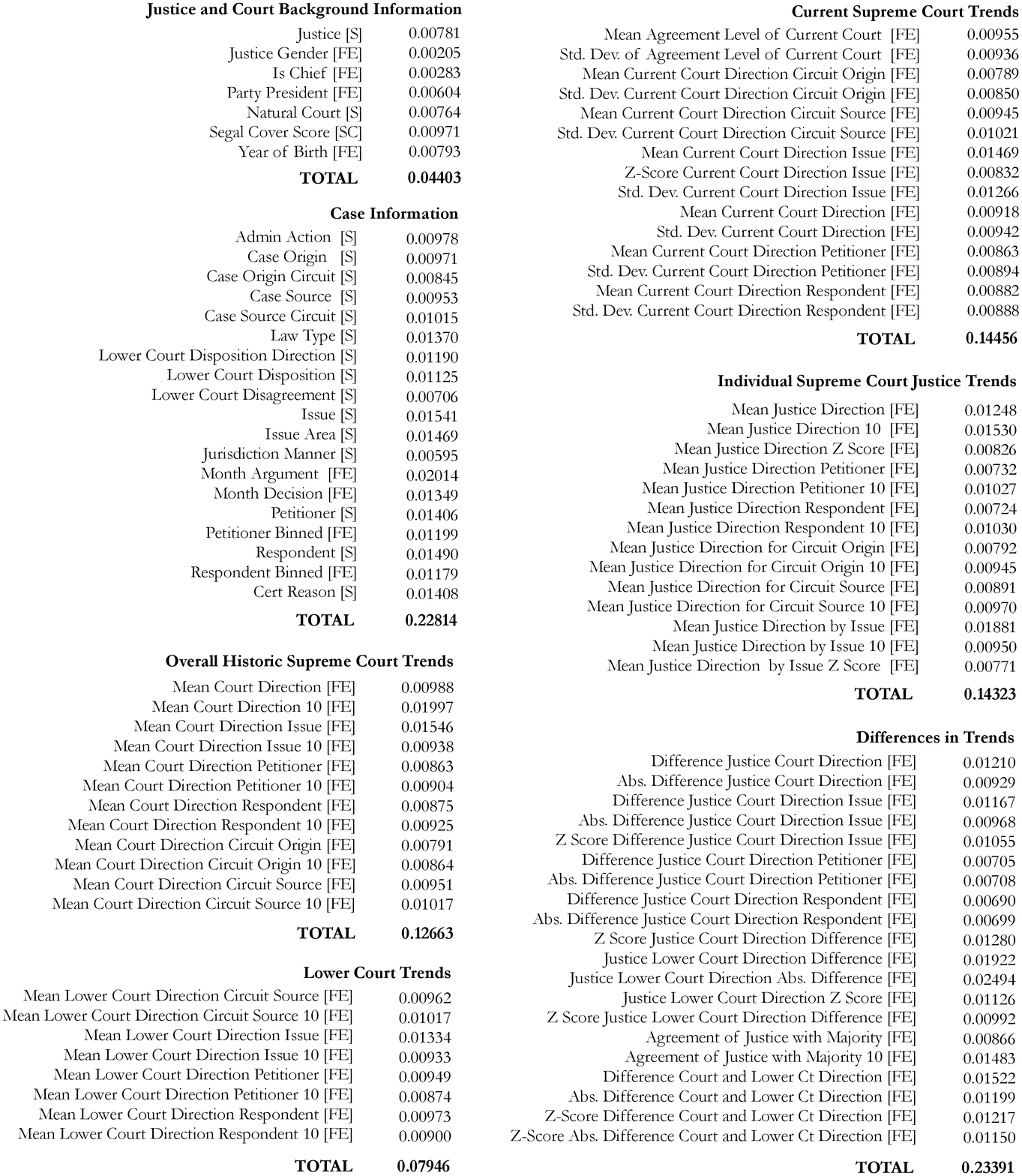}
\caption{Final Feature Weights as of June 2013}
\label{Final Feature Weights as of June 2013}
\end{figure}

\textit{Figure 4} presents the final feature weights as of June 2013.  While these feature weights have adapted over time to take stock of changes in Court's behavior, the June 2013 feature weights provide useful insight into how our model operates.  It is important to note that many of these features are highly correlated.  This ultimately complicates their interpretation (\cite{tolocsi2011classification}, \cite{strobl2008conditional}). Although we present the marginal predictive contribution of each feature, we are mindful of the inherent issues associated with their estimate / interpretation.  That said, the seven larger categories we present in \textit{Figure 4} provide a more reliable (albeit less granular) perspective regarding the contribution of various classes of features.  Collectively, individual case features account for approximately 23\% of predictive power while Justice and Court level background information account for just 4.4\%.  Much of the predictive power of our model is driven by tracking a variety of behavioral trends. This includes tracking the ideological direction of overall voting trends as well as the voting behavior of various justices.  Differences in these trends prove particular useful for prediction.  These include general and issue specific differences between individual justices and the balance of the Court as well as ideological differences between the Supreme Court and lower courts.  
\\
\\\textit{}
Notwithstanding a few notable efforts, \cite{guimera2011justice}, \cite{blackman2011fantasyscotus}, \cite{ruger2004supreme}, and \cite{martin2004competing}, to date, almost all legal and social science scholarship has been backward looking whereby scholars seek to interpret, explain or harmonize prior court behavior.  Even fewer efforts have sought to identify the marginal forward predictive contribution of various case, justice, and other temporal features. We identify features that matter for forward-looking predictions and the relative extent to which they aid in ex ante prediction.  Our approach is also replicable and modular so future scholars can substitute or amend our feature set to determine its impact upon model performance.  
\\
\\  

\section*{Justice and Court Level Temporal Predictability}	
\begin{figure}[ht]\centering
\includegraphics[width=\linewidth]{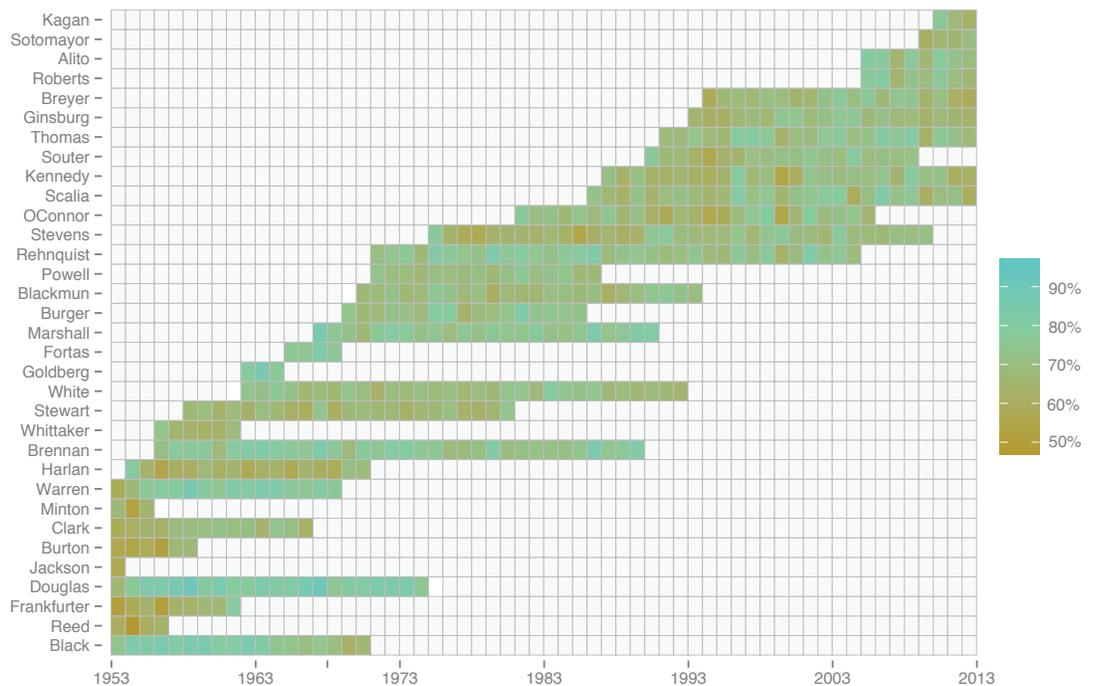}
\caption{Heatmap of Justice and Court Level Temporal Predictability}
\label{Heatmap of Justice and Court Level Temporal Predictability}
\end{figure}

\textit{Figure 5} is heatmap that tracks the temporal performance of our model by Justice. Each column represents a year and each row represents an individual justice.   For each \{justice, year\} pairing corresponding to a year of service, the cell of the heatmap is colored according to the Justice level performance of our model. The more green the cell, the more predictable the Justice in that year. In a few instances where an outgoing and incoming Justice overlapped in a given year the column for that year features with ten colored cells instead of the standard nine. Although we generally err on the side of visualizing all of the data, we do not display cells for Justices that participated in a small fraction of cases in that given term. For example, a Justice who was appointed late in a term, or a Justice who stepped down early in the term. 
\\
\\ 
Reviewing \textit{Figure 5}, our method performs well at predicting certain Justices and not as well on others.  For example, Justices Harlan, Frankfurter, and Burton prove relatively difficult to predict. By contrast, our method is fairly accurate at predicting the behavior of Justices Douglas, Brennan, and Thomas.  This perhaps can be explained by their status as fairly far from the ideological center. 
\\
\\
From an overall perspective, one trend that is immediately apparent is the general level of stability in the quality of our predictions.  Following a short learning period, our ability to predict a given Justice tends to follow a regularized pattern.  In other words, whether we are able to predict a specific case accurately or not, our long run performance is relatively stable for most Justices within a small window of time.  There are, of course, notable exceptions. Justice Stevens begins as a difficult to predict justice but over time becomes increasingly easier to predict.  In his years as an Associate Justice our performance in predicting William Rehnquist is relatively strong.  This changes almost immediately following his elevation to Chief Justice in 1986 when our performance begins to decline.  

\section*{Algorithm Performance By Vote Configuration}	
\begin{figure*}[h]
\centering
\includegraphics[width=\linewidth]{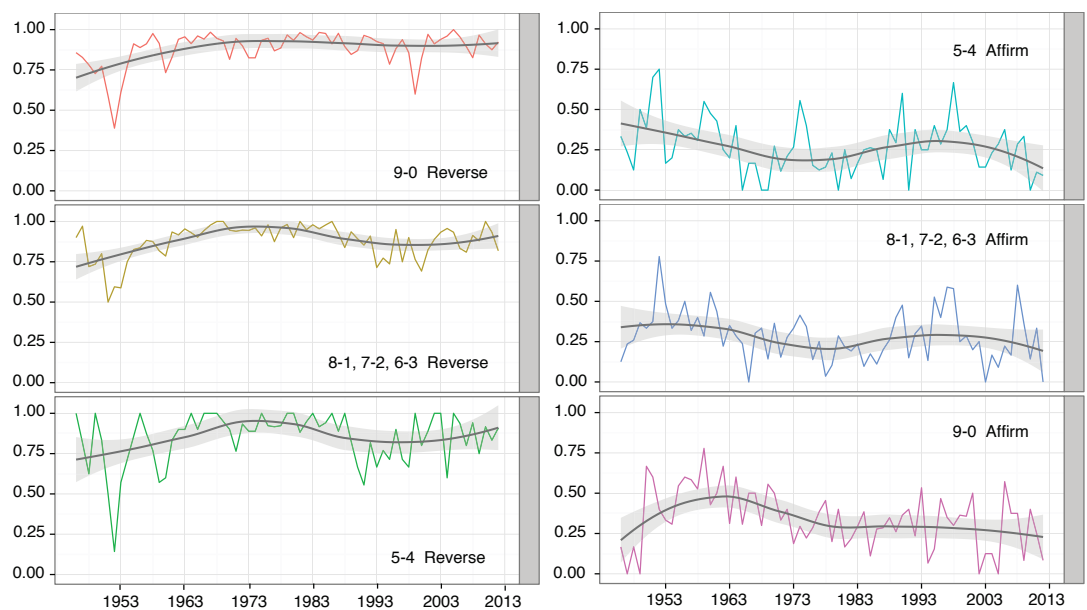}
\caption{Algorithm Performance by Vote Configuration}
\label{testing}
\end{figure*}
\hspace{3 mm}

In addition to exploring the overall and justice level performance of our model, we also seek to explore how the ultimate decision (affirm / reverse) and vote breakdown (9-0 through 5-4) might impact a case's predictability. Our naïve assumption was that our model would perform best on cases where the Court was in agreement and would perform worst in cases with high levels of disagreement among members of the Court.  \textit{Figure 6} supports this basic proposition.  It also once again highlights the strengths and weakness in our model, which performs better for reverse judgments than for affirm judgments.    
\\
\\
With respect to cases that ultimately lead to reversals, our model tracks the commonplace intuition that 9-0 reversals are easier to forecast than 5-4 reversals.  While our performance between these categories is somewhat close in certain years, we consistently perform better in unanimous reversal cases than in cases which feature disagreement between justices. We also perform better on cases with a vote of 9-0 to affirm than in cases that affirm through a divided court.  However, as demonstrated earlier, it is clear that our model struggles to identify in advance cases that the Court ultimately decides to affirm.  Since 1953, the Court has affirmed 2,623 cases or 34.1\% of its fully argued cases.  On this subset of cases, our model does not perform particularly well.  In some years, we are able to forecast less than 25\% of these cases correctly.  
\\
\\
An open question for future research is whether this failure to accurately predict affirmed decisions is a permanent fixture of underlying stochasticity in Court’s behavior or a failure of the model or available feature set to identify some latent dimension that would allow for better forecasting of the reversal / affirm decision. 

\section*{Conclusion}

Using only information known prior to the Court's decision, case by case and term by term, we construct a model that predicts each decision of the Supreme Court of the United States from 1953 - 2013. Leveraging extremely randomized trees, a particular form of ensemble tree model, we correctly forecast 69.7\% of the Court’s overall affirm / reverse decisions and  70.9\% of the votes of individual justices across the 7,700 cases and more than 68,000 justice votes.  We offer a major contribution to the science of quantitative legal prediction by generating the first general, robust and fully predictive model of Supreme Court decision making offered to date.

\bibliography{main}

\end{document}